\documentclass[rotating]{aa}
\usepackage{amsmath}                               
\usepackage{txfonts}
\usepackage[final]{epsfig} 
\usepackage[final]{graphics}

\usepackage{natbib}
\usepackage[english,american]{babel}

\begin{document}
\def\first{1$\rm^{st}$~}
\def\second{2$\rm^{nd}$~}
\def\third{3$\rm^{rd}$~}
\def\fourth{4$\rm^{th}$~}
\def\irascolor{$\Delta I_{60}$/$\Delta I_{100}$}
\title{Far-infrared loops in the \second Galactic Quadrant}

\author{ Cs.~Kiss\inst{1,2}
  \and  A.~Mo\'or\inst{1} 
  \and  L.~V.~T\'oth\inst{2,3}   }

\institute{
  Konkoly Observatory of the Hungarian Academy of Sciences, 
    P.O. Box 67, H-1525~Budapest, Hungary
  \and Max-Planck-Institute f\"ur Astronomie, K\"onigstuhl 17, 
    D-69117, Heidelberg, Germany 
  \and Department of Astronomy, E\"otv\"os Loránd University, 
    P.O. Box 32, H-1518~Budapest, Hungary}
\offprints{Cs.~Kiss, pkisscs@konkoly.hu}

\date{ Received  / Accepted ...}

\abstract{We present the results of an investigation
of the large scale structure of the diffuse interstellar medium 
in the \second Galactic Quadrant (90\degr\,$\le$\,l\,$\le$\,180\degr). 
145 loops were identified on IRAS based far-infrared maps.
Our catalogue lists their basic physical properties.
The distribution clearly suggests that there is an efficient
process that can generate loop-like features at high galactic latitudes. 
Distances are provided for 30 loops. 
We also give an observational estimate on the volume filling factor
of the hot gas in the Local Arm, 
4.6\%\,$\le$\,$f_{2^{nd}}$\,$<$\,6.4\%. 
}
	  
\maketitle

\section{Introduction \label{sect:introduction}}

A study of the diffuse interstellar medium is presented, a step in the way 
deriving the near-past and near-future of the galactic star formation.
The large scale structure of the ISM in the Galaxy is diverse with 
the complex distribution of cavities, filaments, arc, loops and shells,
which is often referred to as the ``Cosmic Bubble Bath'' 
\citep{Brand}. Ridges of enhanced radio contiunuum emission 
{ extending} to high galactic
latitudes have been known since the 60s. 
The first extensively studied examples are Loop\,I
\citep{Large66}, Loop\,II \citep["Cetus Arc"][]{Large62} and
Loop\,III \citep{Large62}. 
Later anomalous HI features like worms, chimneys, supershells were identified
and catalogued, up to a characteristic size of $\sim$2\,kpc with 
expansion velocities 
of $\sim$20\,kms$^{-1}$ \citep{Heiles79,Heiles80,Heiles84,Hu,Koo91}. 
These were associated with voids in the HI distribution 
\citep[see e.g.][for a review]{Palous96}.  
Similar complex morphology has been recognized in some nearby spiral and
irregular galaxies e.g. M\,31 \citep{Brinks}, M\,33 
\citep{Deul} and the Magellanic Clouds 
\citep{Staveley,Kim}. 


The most studied galactic shells were formed by 
supernova (SN) explosions and winds of massive stars \citep{TT}. 
However, high velocity clouds (HVCs) may also form large cavities
when infalling from the galactic halo and colliding with the ISM of the disc. 
A remarkable example of a HVC -- galactic disc interaction
is the North Celestial Pole (NCP) loop \citep{Meyerdierks}. 
\citet{Ehlerova} investigated the origin of galactic HI
shells and found, that those are likely related to star formation,
and not to the infall of HVCs. 
Apart from the connection of these features to young massive stars and their
associations, they may also provide a nest for the next generation of stars
via the process of propagating star formation.
The idea was first proposed by \citet{Oort} and \citet{Opik}, and
later followed by observational evidences \citep[see][for a review]{Blaauw}.
{ This mechanism was modelled in detail by \citet{Elmegreen+Lada} and
successfully applied to the sequential formation of subgroups in OB
associations}.  
Gamma ray bursts may also be responsible for the formation of the
largest HI holes, as proposed by \citet{Efremov} and \citet{                  Loeb}.
Nonlinear development of instabilities in the ISM may form 
large cavities as well, without stellar energy injection
\citep{Wada,Korpi,Klessen,Sanchez}.

The cavities of the ISM contain low density hot gas in the multiphase model
by \citet{McKee}. The filling factor $f$ of this hot gas in the galactic disc is
an important parameter of the ISM, which is poorly constrained observationally.
\citet{Ferrier98} and \citet{Gazol}
estimate $f$\,$\approx$\,20\% at the solar circle, which drops 
to $f$\,=\,3--8\% for  
8.5\,$\le$\,$R_{\odot}$\,$\le$\,10\,kpc galactocentric distances.

The 21\,cm emission of diffuse HI clouds is well-correlated with the 100\,$\mu$m
FIR emission \citep{Boulanger+Perault}. 
This extended emission is often referred to as the galactic cirrus 
\citep{Low}. A typical far-infrared colour of
$\Delta I_{60}/\Delta I_{100}$\,=\,0.25 was derived from IRAS/ISSA maps
when the gains are corrected 
according to \citet{Wheelock}. According to COBE/DIRBE results     
its dust colour temperature is T$_d$\,$\approx$\,18\,K 
\citep{Lagache98}.

Studies of the large scale structure of the cirrus emission and 
loop- or arc-like features were restricted so far
to low galactic latitudes in the far-infrared 
\citep{Schwartz87,Marston}, and reached only medium 
galactic latitudes in radio surveys. 
Detailed investigations
of individual loops were performed only in the vicinity of the galactic plane.
Recently \citet{Daigle} presented the first automatic tool 
for detecting expanding HI shells using artificial neural networks.
Their data is restricted to the sky area 75\degr\,$\le$\,l\,$\le$\,145\degr and
$-$3\degr\,$\le$\,b\,$\le$\,5\degr and their detection strategy is based on
the velocity information of the HI data rather than morphology.  
 
In order to get a complete, full-scale view on the large-scale structure
of the cold diffuse ISM, in our study
we searched the whole \second galactic quadrant 
(90\degr\,$\le$\,l\,$\le$\,180\degr; --90\degr\,$\le$\,b\,$\le$\,90\degr) 
for intensity enhancements in the
interstellar medium, showing loop-like structures.

\section{Data analysis \label{sect:data}}

\subsection{Quest for loops on far-infrared maps \label{search}}

We investigated the 60 and 100\,$\mu$m ISSA plates 
\citep[IRAS Sky Survey Atlas,][]{Wheelock} 
in order to explore the distribution of dust emission in 
the \second Galactic Quadrant. 
We created composite images of the 12{\fdg}5$\times$12{\fdg}5 sized 
individual ISSA plates using the "geom" and "mosaic" procedures
of the IPAC-Skyview package, both at 60 and 100\,$\mu$m. 
These images were built up typically 
from $\sim$10-15 ISSA plates, reaching a size of 
$\sim$\,40{\degr}$\times$40{\degr}.
Loop-like intensity enhancements were searched by eye on the 100\,$\mu$m 
mosaic maps. Loops by our definition must show an excess 
FIR intensity confined to an arc-like feature, 
at least 60\% of a complete ellipse-shaped ring. 
A loop may consist of a set of bright, more or less isolated, 
extended spots, or may be a diffuse ring or part of a ring. 
The size of the mosaic image limits the maximal size of the objects found.
On the other hand, due to the relatively large 
size of the investigated regions, loop-like intensity enhancements 
with a size of $\le$\,1\,\degr were not searched. 
The original ISSA $\rm I_{60}^{ISSA}$ and 
$\rm I_{100}^{ISSA}$ surface brightness values were transformed 
to the COBE/DIRBE photometric system, using the 
conversion coefficients provided by \citet{Wheelock}: 
\begin{itemize}
\item $\rm I_{60} = 0.87\times I_{60}^{ISSA} + 0.13\,MJysr^{-1}$ 
\item $\rm I_{100} = 0.72\times I_{100}^{ISSA} - 1.47\,MJysr^{-1}$
\end{itemize}

Dust IR emission maps by \cite{Schlegel} (SFD) were investigated 
to derive parameters describing our loop  features (see Sect.~2.2).
The main differences of the SFD 100\,$\mu$m map compared to the ISSA maps 
are the following: 
\begin{itemize}
\item[(1)] Fourier-destripping was applied, 
\item[(2)] asteroids and non-Gaussian noise were removed,
\item[(3)] IRAS and DIRBE 100\,$\mu$m maps were combined, preserving the
    DIRBE zero point and calibration,
\item[(4)] stars and galaxies were removed.
\end{itemize}
We analysed the radial surface brightness profiles of the loops on the 
SFD 100\,$\mu$m map in order to check the effect of the removal
of the sources mentioned above.   
We also used the SFD $\rm E(B-V)$ maps derived from the dust column density 
maps. In the case of these maps the colour temperature was derived from 
the DIRBE 100 and 240\,$\mu$m maps, and a temperature corrected map
was used to convert the 100\,$\mu$m cirrus map to a map proportional
to dust column density.


\paragraph{Shape:}
We approximated the shape of a possible loop by an
ellipse, which was then fitted using a 2D least-square fit method.
An ellipse shape is expected from SN or stellar wind shells, 
since (1) non-spherical explosion (wind) may occur, in the 
most extreme case creating a ring, rather than a shell, and
(2) originally spherical shells are distorted to ellipsoidal 
(i) due to the shear in the direction of galactic 
rotation \citep{Palous} and 
(ii) due to the vertical gravitational field in the galactic disc
\citep[see e.g.][]{Ehlerova}.  
  
\paragraph{The fitted ellipse} is defined with
the central (galactic) 
coordinates, the minor and major semi-axis of the ellipse, and
the position angle of the major axis to the circle of galactic 
latitude at the centre of the ellipse. This latter was defined 
to be '+' from East to North (or counter-clockwise).  

\begin{figure}
\centerline{\hbox{\epsfxsize=8.5cm \epsffile{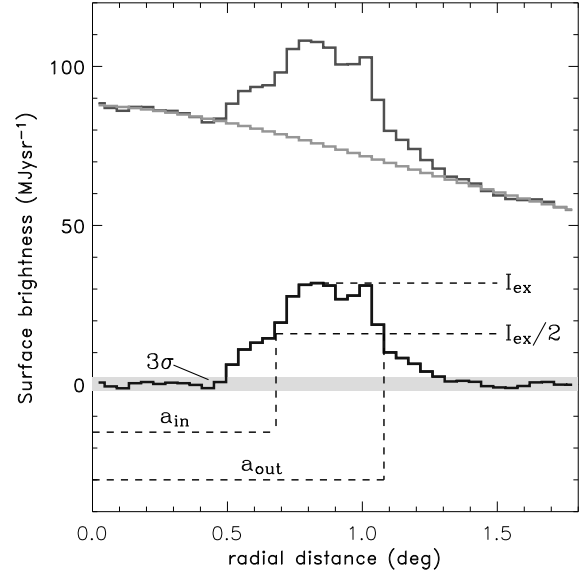}}}
\caption{Intensity profile before (top) and after (bottom) background removal
(gray solid line) with the main derived parameters
($I_{ex}$, $\sigma_{ex}$, $a_{in}$, $a_{out}$, see Sect.~2.2)}
\label{fig:q}
\end{figure}

\paragraph{Intensity profile:} 
For all of our loops we extracted radially averaged surface
brightness profiles, extended to a distance of 
twice the major (and minor) axis of the fitted ellipse,
using 40 concentric ellipsoidal rings. 
These surface-brightness profiles (ISSA 100 and 60\,$\mu$m, SFD 100\,$\mu$m
and SFD reddening maps) 
were used in the following to determine the basic parameters of the
FIR emission in the loop. An example is shown in Fig.~\ref{fig:q} 


\begin{figure*}
\hbox{\epsfig{file=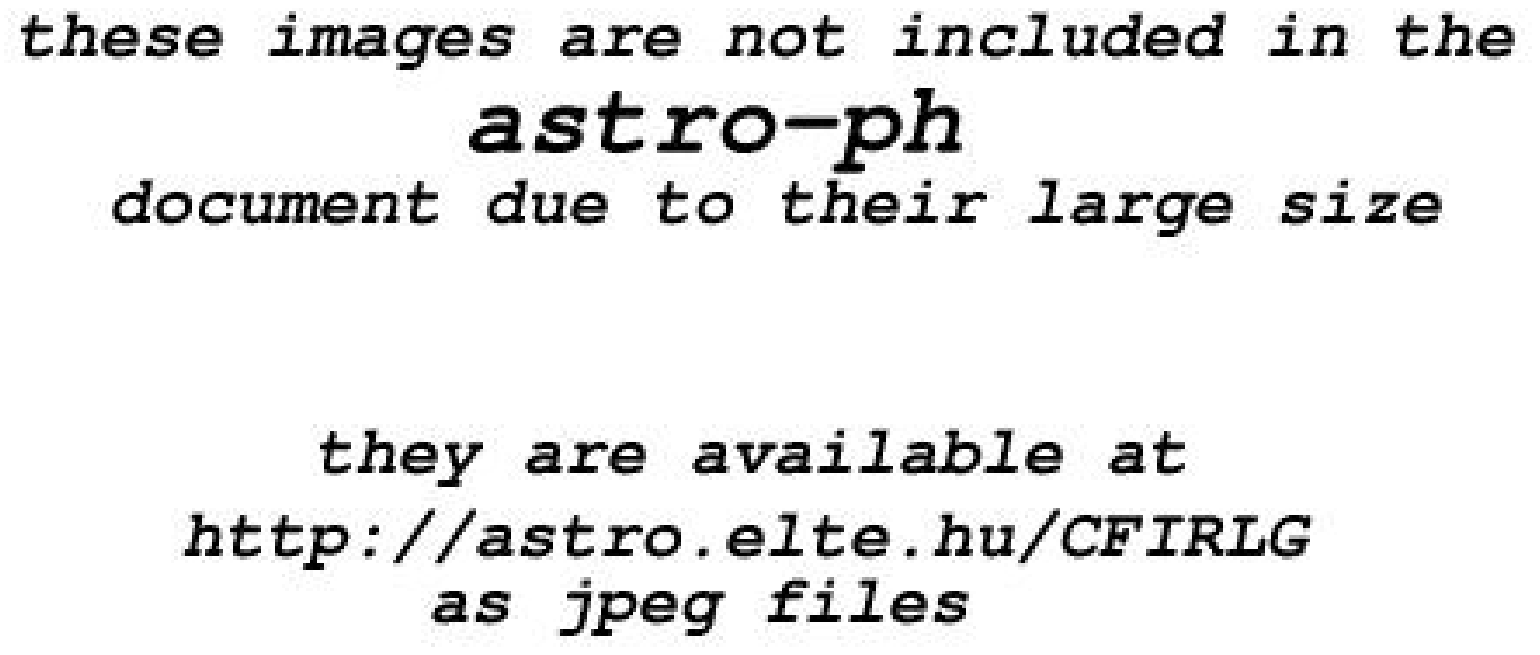, width=14cm} 
}
\caption{Examples of loops found in the \second Galactic Quadrant. 
{\it left:} GIRL~G134-- 28, first described by \citet{Heiles84} as 
\object{GS\,137--27--17}. This supershell is located at a distance
at 1.5\,kpc and has a size of $\sim$0.41$\times$0.31\,kpc; 
{\it right:} GIRL~G117+00, supershell around the Cas\,OB5 association
in a distance of 2.5\,kpc with a size of $\sim$0.31$\times$0.27\,kpc
\citep{Moor}. 
}
\label{fig:images}
\end{figure*}

\subsection{Derived parameters \label{sect:derived_params}}

\paragraph{Significance:}
The local background was determined using the 'non-loop' points in the 
radial surface brightness
profiles, fitting a 3rd order polynomial. This appropriate 
background was removed from each surface brightness profile points. 
The intensity excess $I_{ex}$ was derived as the maximum value of this 
background removed profile. In order to check if this value is 
above the 'noise', we calculated the standard deviation of the background
removed intensity in the 'non-loop' values, $\sigma_{ex}$,
and defined the significance of the loop as $\Psi = I_{ex}/ \sigma_{ex}$. 
We derived significance parameters on 60 and 100\,$\mu$m ISSA maps and
on the SFD 100\,$\mu$m point source removed sky brightness and reddening maps.
The higher the value of $\rm \Psi$ is the higher the intensity 
excess of the loop over the background, therefore we use
this parameter as a 'quality indicator' in the following.

\paragraph{Relative width:}
Inner and outer edges of the loop wall along the major axis 
($a_{in}$ and $a_{out}$, respectively) are defined as the radial distance
at the full width at half power of the
background removed intensity profile, $I_{ex}/2$ (Fig.~\ref{fig:q})
We distinguish three regions for a specific loop:
(1) loop interior (a\,$<$\,a$_{in}$), 
(2) loop wall (a$_{in}$\,$\le$\,a\,$\le$\,a$_{out}$), 
(3) outer region (a\,$>$\,a$_{out}$).
The relative width of the wall of the fitted ellipse is defined as
$W\,=\,1-a_{in}/a_{out}$. 

\paragraph{Colour index:}
We derived a colour index for our loops $\Delta I_{60}$/$\Delta I_{100}$, 
from the radially averaged 60 and 100\,$\mu$m surface brightness
profile. This is defined as the slope of the $\rm I_{60}$ vs. $\rm I_{100}$ 
scatter plot using the data points of the 
surface brightness profiles in the positions of the loop wall 
($a_{in}$\,$\le$\,$a$\,$\le$\,$a_{out}$) only.


\section{Results}

\subsection{The catalogue \label{sect:catalogue}}
We identified 145 loops in the 2nd Galactic Quadrant. 
The parameters of these loops are 
summarized in Table~A.1. 
We call these objects 
'{\sl GIRL}'-s, abbreviating 
'{\sl G}alactic {\sl I}nfra{\sl R}ed {\sl L}oops'.
\begin{figure}
\epsfig{file=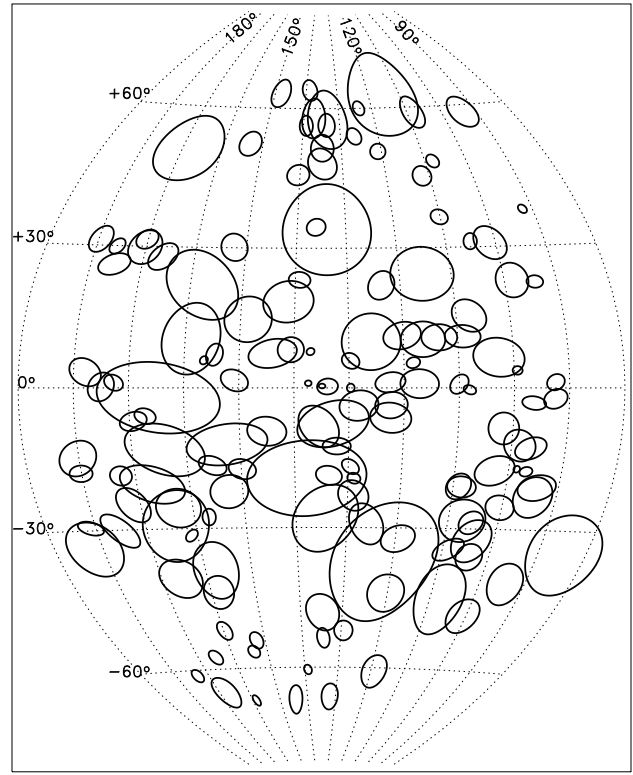, width=8.5cm}
%
\caption[]{Distribution of GIRLs in the sky 
in the \second Galactic Quadrant (Aitoff projection), 
represented by the fitted ellipses}  
\label{fig:mainsky}
\end{figure}

The entries of the catalogue are the following: 
\begin{itemize}
    \item[1.)] Name of the loop, derived from the galactic coordinates of the 
             loop centre. The format is : 
	     GIRL~GLLL$\pm$BB, the 'GIRL'  
	     prefix stands for 'Galactic InfraRed Loops', 'G' 
	     marks the galactic coordinate system, 'LLL' is the
	     galactic longitude in degrees, $\pm$ is the sign of the galactic
	     latitude (+/--), and 'BB' is the absolute value of the 
             galactic latitude of the loop centre.
    \item[2.)] Central galactic coordinates of the loop (l and b)
    \item[3.)] Semi major and minor axis of the fitted ellipse
    \item[4.)] Position angle of the fitted ellipse. The position angle
             is defined as '+' from East to North. Zero position angle
	     is pointed to the East in galactic coordinates.
    \item[5.)] Major axes corresponding to the inner and outer edge of 
              the loop wall, 
             derived from the 100\,$\mu$m surface brightness profile
    \item[6.)] { ISSA 60 and 100\,$\mu$m, SFD 100\,$\mu$m and reddening 
    	significance parameters ($\Psi_{I60}$, $\Psi_{100}$, 
	 $\Psi_{S100}$ and $\Psi_{EBV}$, respectively}
    \item[7.)] colour index  $\Delta I_{60}$/$\Delta I_{100}$ of 
    		the loop wall 
\end{itemize}   		 
In the electronic version of the catalogue (URL: ``http://astro.elte.hu/CFIRLG'') 
we provide the following additional features:
\begin{itemize}
    \item[8.)] 100\,$\mu$m ISSA image of the loop with the possible associated
             objects overlaid
    \item[9.)] { ISSA 60 and 100\,$\mu$m, SFD 100\,$\mu$m and
    	SFD E(B-V) surface brightness profiles in FITS format}
    \item[10.)] List of objects apparently associated with the loop
\end{itemize}

An associated object has to be placed in the wall or in the 
interior of the loop (defined by concentric ellipses, as described above).
We considered the following type of possible associated objects 
(references are indicated)

\begin{itemize}
    \item[--] dark clouds \citep{Dutra}
    \item[--] supernova remnants \citep{Green} 
    \item[--] OB-associations \citep{Lang}
    \item[--] pulsars \citep{Taylor}
    \item[--] HII regions \citep{Sharpless}  
    \item[--] IRAS point source with molecular core FIR colours
    \item[--] IRAS point source with T Tau star-like FIR colours 

\end{itemize}
{ 
Molecular cores and T~Tauri stars were selected from the IRAS Point Source
Catalogue (Joint IRAS Science Worknig Group, 1988) according to the
following criteria:
\begin{itemize}
\item point sources associated with galaxies were excluded
\item photometric qualities are 2 or better at 12, 25 and 60\,$\mu$m
\item \begin{itemize}
           \item molecular cores: \\
           0.4\,$\le$\,$log_{10}\bigg( {{F_{25}}\over{F_{12}}} \bigg)$\,$\le$\,1.0
           \& 
           0.4\,$\le$\,$log_{10}\bigg( {{F_{60}}\over{F_{25}}} \bigg)$\,$\le$\,1.3
        \item T Tauri stars: \\
           0.0\,$\le$\,$log_{10}\bigg( {{F_{25}}\over{F_{12}}} \bigg)$\,$\le$\,0.5
           \& 
           --0.2\,$\le$\,$log_{10}\bigg( {{F_{60}}\over{F_{25}}} \bigg)$\,$\le$\,0.4         
       \end{itemize}  
       following the definitions by Emerson~(1998). F$_{12}$, F$_{25}$ and 
       F$_{60}$ are the 12, 25 and 60\,$\mu$m uncorrected IRAS fluxes, 
       respectively. 
\end{itemize}
}

This list of possible associated object does not take into account
the distances of the individual objects, therefore all objects
projected to the loop wall or to the interior are included.
In the case of a few (30) loops, we were able to derive a distance 
for the loop based on the distances of associated objects.
This is discussed in detail in Sect.~5. 
{ The list of associated objects is presented in Appendix~C}.


\begin{figure}

\epsfig{file=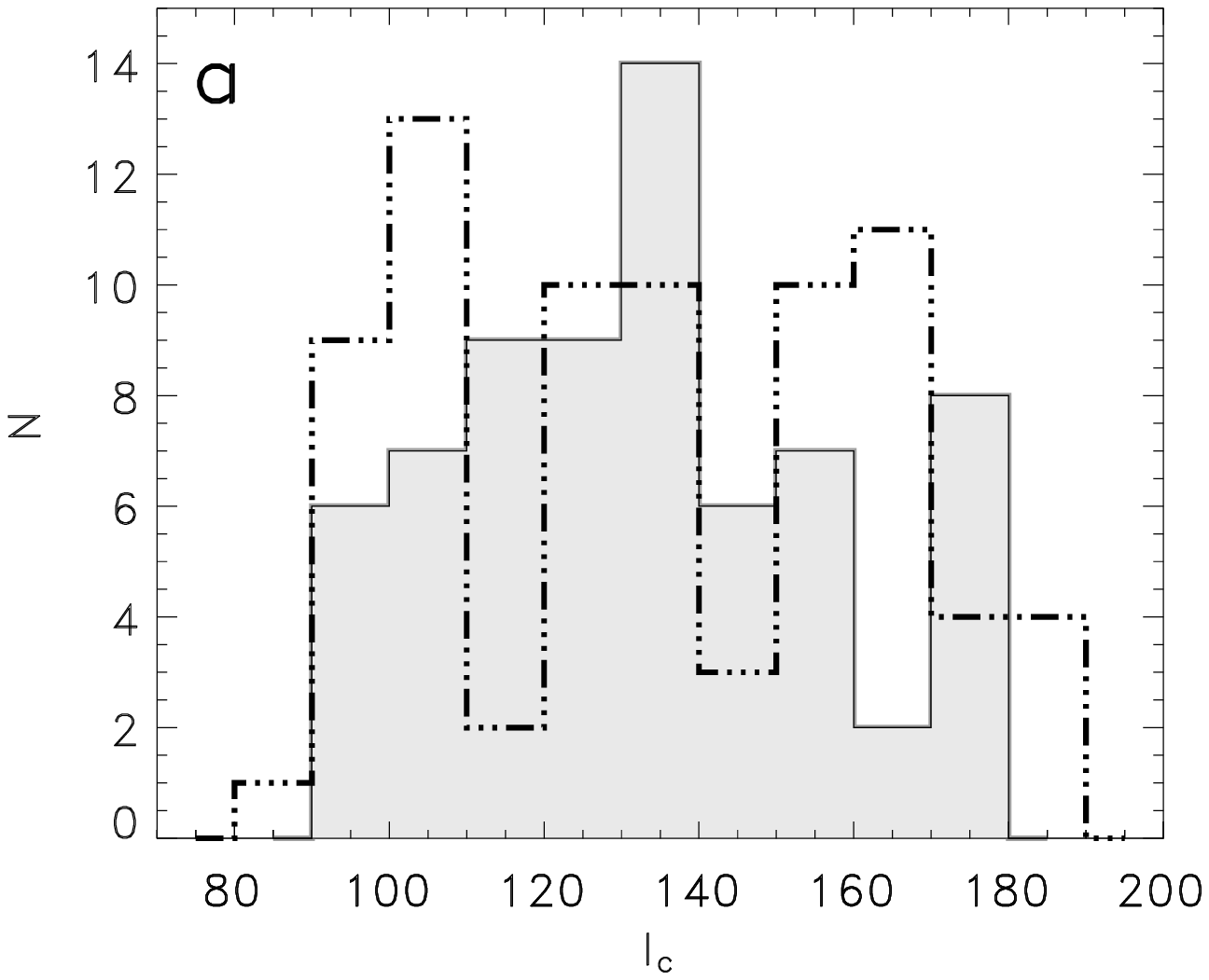, width=8.5cm}
\\
\epsfig{file=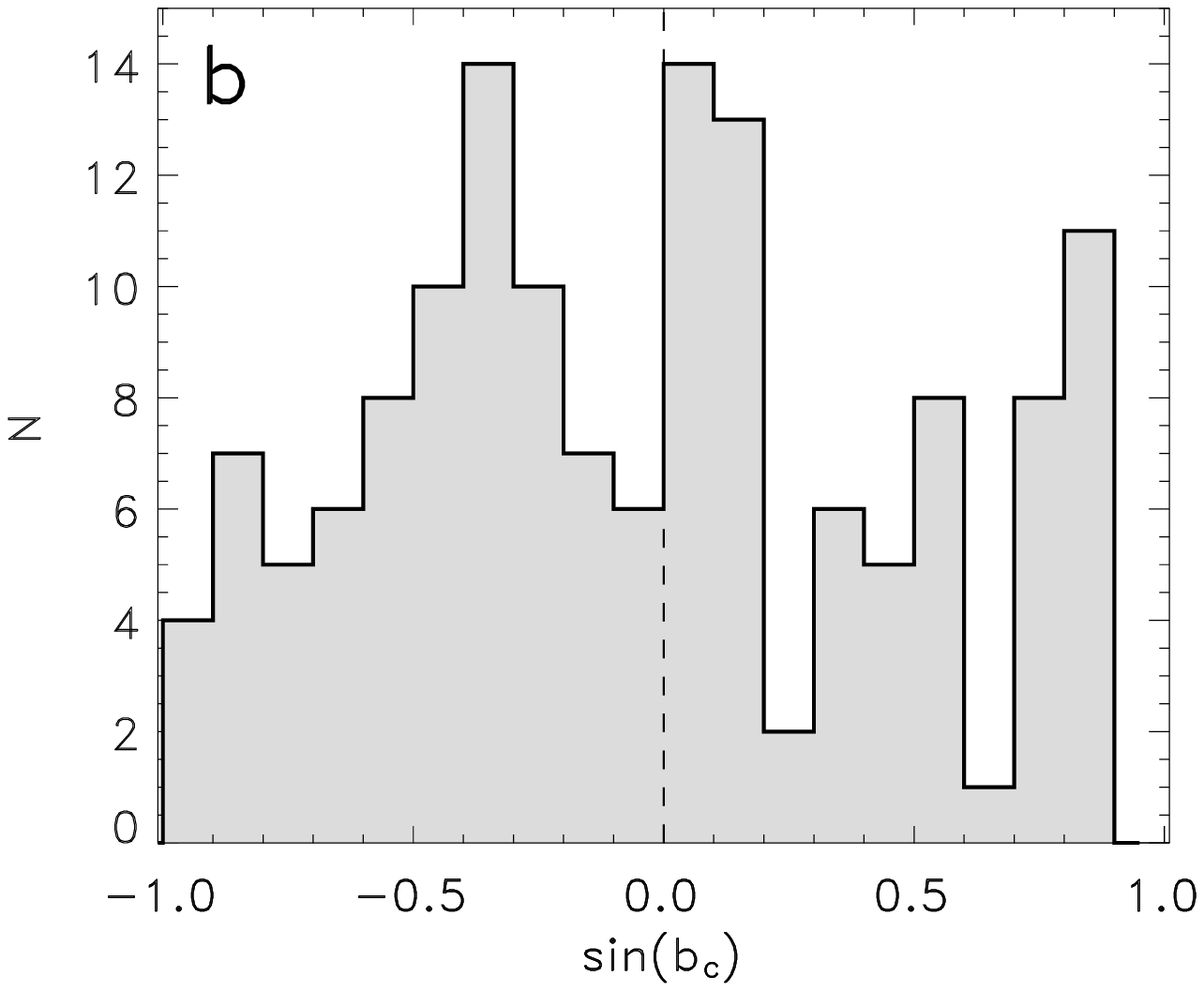, width=8.5cm}
\caption[]{Distribution of GIRLs in the 2nd Galactic Quadrant.
({\bf a}) Loop-counts vs. galactic longitude. Gray lines with 
  light gray filled area: northern galactic hemisphere (b$>$0\degr). 
  Dash-dotted black lines: southern galactic hemisphere (b$<$0\degr).
({\bf b}) Loop-counts vs. the 'sin' of the galactic latitude (equal area belts)}  
\label{fig:mainhistogram}
\end{figure}


\subsection{Sky distribution and fundamental statistics}

\paragraph{Sky distribution:}
Fig.~\ref{fig:mainsky} presents the distribution of the far-infrared loops
found in the 2nd Galactic Quadrant.
The distribution of GIRLs seems to be rather complex.
However, there are slightly more loops at low (b\,$\le$\,30\degr) 
galactic latitudes than at high ones (b$>$30\degr).      

It is obvious from Fig~\ref{fig:mainhistogram}a 
that the two galactic 
hemispheres represent different loop distributions.
We have tested the hypothesis that the galactic latitudes of the loop centres 
on the Northern and Southern hemispheres are drawn from the same distribution.  
A Kolmogorov--Smirnov test showed that the homogeneous distribution hypothesis 
could not be accepted at the 80\% confidence level. 
A test performed for the galactic longitude 
distribution on the two hemispheres (Fig.\ref{fig:mainhistogram}b) yielded  
the same result.

There is a hint in { Fig.~\ref{fig:mainsky}}, 
that most of the loops are part of a structure 
with scales significantly larger than the individual loop diameters. 
The most remarkable
feature of this kind is a huge chain of loops seen in the 
l\,$\approx$\,90\degr--130\degr,  b=--35\degr--\,+5\degr region. 
In the ranges l\,$\approx$\,130\degr--140\degr, 
b\,$\approx$\,+45\degr-- +65\degr nine loops can 
be found squeezed to a relatively small area, which correlates well with 
the position of the Ursa Maior molecular complex
connected to the expanding shell of the NCP loop \citep{Pound}.

\paragraph{Size distribution:}
\begin{figure}[h!]
\epsfig{file=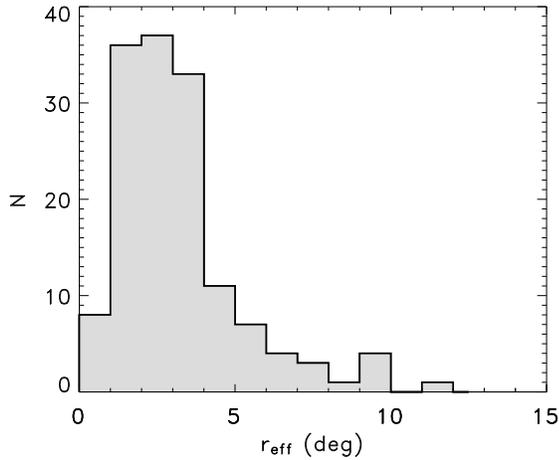, width=8.5cm}
\caption[]{Apparent size distribution of GIRLs. Size is presented in 
effective radius, r$\rm_{eff}$, the geometrical mean of semi-minor and 
-major axes of the fitted ellipse}
\label{fig:size}
\end{figure}
We defined the effective radius of our loops as the
geometrical mean of the semi-minor and -major axis of the fitted ellipse,
$r_{eff} = \sqrt{a\times b}$.
For the whole sample of loops we obtained 
$ m(r_{eff}) = 3\fdg2$ and 
$E(r_{eff}) = 2\fdg7$ for the 
{ sample mean and expectation value} of the 
effective radius, respectively. 

The size distribution is presented in Fig.~\ref{fig:size}.
Majority (78\%) of the loops are in the apparent effective radius 
$r_{eff}$\,$\le$\,4\degr, and there are only a few (9) loops 
$r_{eff}$\,$\ge$\,7\degr. 
The lower end of the 
size distribution is limited by our detection method, which excludes features 
with a size smaller than $\sim$1\degr in diameter.

\paragraph{Distribution of relative width:}   
\begin{figure}[h!]
\epsfig{file=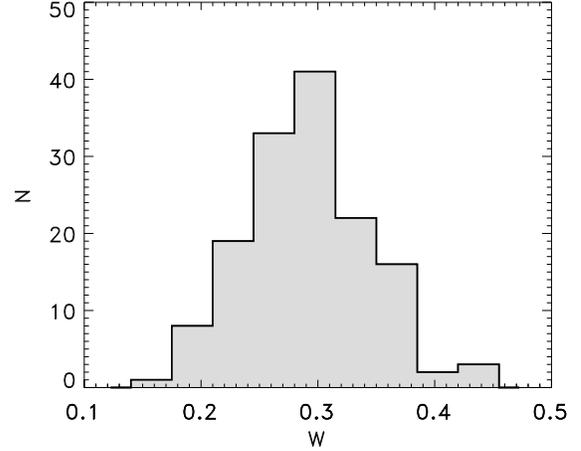, width=8.5cm}
\caption[]{Distribution of the width of the loop walls relative to 
the effective radius of the loop (W = 1 -- a$_{in}$/a$_{out}$)}
\label{fig:thickness}
\end{figure}

Fig.~\ref{fig:thickness} shows a Gaussian-like distribution, 
with an average value of $\rm \langle W \rangle$\,=\,0.29 and a 
FWHM\,=\,0.04 for the relative width $W$ of the loop walls.

\begin{figure}[h!]
\epsfig{file=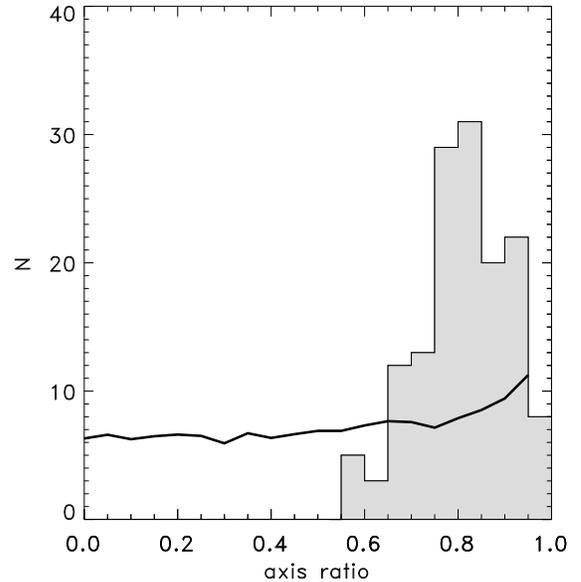, width=8.5cm}
\caption[]{Comparison of the expected distribution of 
minor to major axis ratio (assuming randomly oriented 2D rings, 
solid line) with the distribution of the observed axis ratios
(gray histogram). 
\label{fig:abratio}
}
\end{figure}

\paragraph{Axis ratio distribution:}
As mentioned in Sect.~1, it is widely accepted, that galactic 
loops -- identified in any wavelength range -- are 2D projections
of 3D cavities in the ISM. 
We compared the expected distribution of the ratio of minor to major axes 
of the sky-projection of 2D ring to
the distribution of the observed minor to major axes
ratios in Fig.~\ref{fig:abratio}. 2D rings are assumed to 
be randomly oriented in $\phi$ and $\theta$ polar angles.
This figure clearly shows a significant difference between the 
two distributions. It is not likely that most
of our objects are just the projections of 2D rings. 
  


\paragraph{Significances:}
In Fig.~\ref{fig:significances} we compare the significances derived 
from ISSA 100\,$\mu$m,
ISSA 60\,$\mu$m,  SFD 100\,$\mu$m and SFD reddening maps, 
$\Psi_{I100}$, $\Psi_{I60}$, $\Psi_{SFD100}$, $\Psi_{EBV}$ 
(see also Table~A.1). 
The Pearson correlation coefficients are listed in 
Table~\ref{table:psicoeffs}.
\begin{table}[ht!]
\begin{tabular}{lr}
\hline
correlated & correlation \\ 
significances&   coefficients \\ \hline
$\Psi_{I100}$ -- $\Psi_{I60}$ & 0.77 \\
$\Psi_{I100}$ -- $\Psi_{S100}$ & 0.82 \\
$\Psi_{I100}$ -- $\Psi_{SEBV}$ & 0.68 \\
$\Psi_{S100}$ -- $\Psi_{SEBV}$ & 0.79 \\ \hline
\end{tabular}
\caption[]{Linear Pearson correlation coefficients of significances 
derived from ISSA 100\,$\mu$m ($\Psi_{I100}$), ISSA 60\,$\mu$m ($\Psi_{I60}$), 
SFD 100\,$\mu$m ($\Psi_{S100}$) and SFD E(B--V) ($\Psi_{SEBV}$) maps.}
\label{table:psicoeffs}
\end{table}
The $\Psi_{I100}$, $\Psi_{I60}$, $\Psi_{S100}$ and $\Psi_{SEBV}$
significances are well correlated, and in all 145 cases resulted in a 
definite confirmation of the suspected ellipsoidal structure.  

\begin{figure}
\epsfig{file=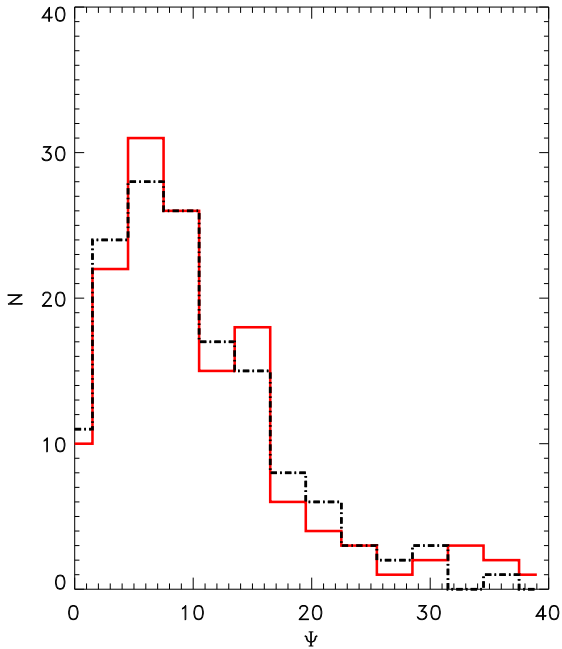, width=8.5cm}
\caption[]{{ Distribution of the $\Psi_{I100}$ (gray solid line)
and $\Psi_{EBV}$ (black dash-dotted line) significance values}}
\label{fig:psihist}
\end{figure}
{
Fig.~\ref{fig:psihist} presents the distribution of ISSA 100\,$\mu$m
and SFD E(B-V) significances ($\Psi_{I100}$ and $\Psi_{EBV}$, respectively).
Based on the definition of $\Psi$ (see Sect~2.2) one can set a limit of 
$\Psi$\,=\,3 as a value separating \emph{weak} and \emph{prominent} loops
(prominent loops are above the ``3$\sigma$'' level of background fluctuations).
However, as seen in Fig.~\ref{fig:psihist},
the majority of our loops show high 
sigma values, regardless of the significancy type ($\Psi_{I100}$, $\Psi_{I60}$,
$\Psi_{S100}$ or $\Psi_{EBV}$) and only very few loops have $\Psi$\,$\le$\,3.

Bright fore- or background sources not related 
to the loop may cause low significances.
An example is GIRL~G125+09, which has low $\Psi_{I100}$ and $\Psi_{S100}$
(2.4 and 2.6, respectively) but it is outstanding on the reddening maps
($\Psi_{EBV}$\,=\,6.0) and is very conspicuous at 60\,$\mu$m 
($\Psi_{I60}$\,=\,17.2). In addition, this loop is clearly detected in the
HI 21\,cm line \citep{Kiss2000}. The main reason behind the low significances
at 100\,$\mu$m is the presence of an extremely bright region in the 
south-eastern part of the loop related to \object{NGC\,7822}, \object{S\,171}
and to the \object{Cep\,OB4} association, which dominates the background of 
GIRL~G125+09.  
} 
%
\begin{figure}
\epsfig{file=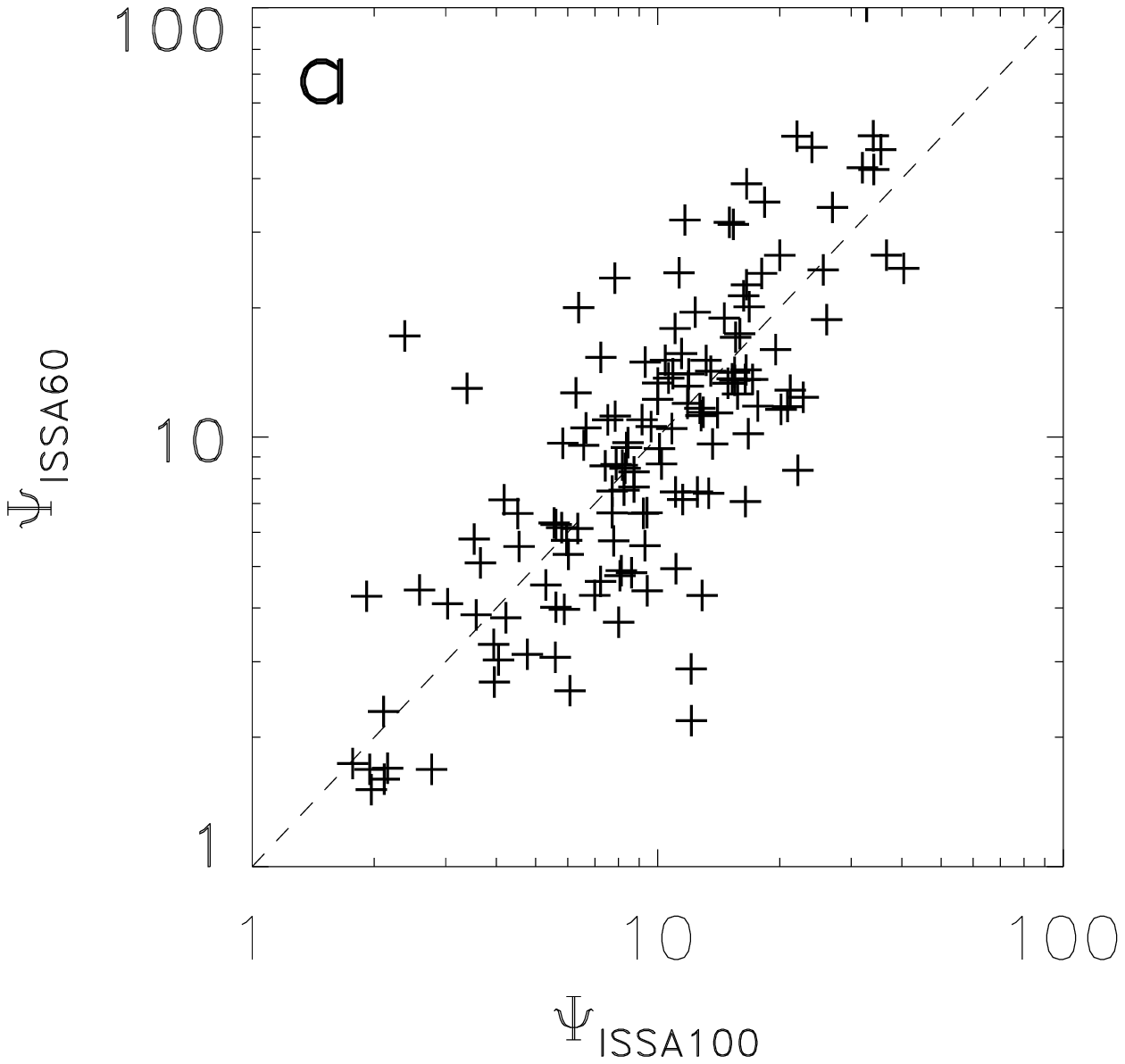, width=4.5cm}
\hskip -0.4cm \epsfig{file=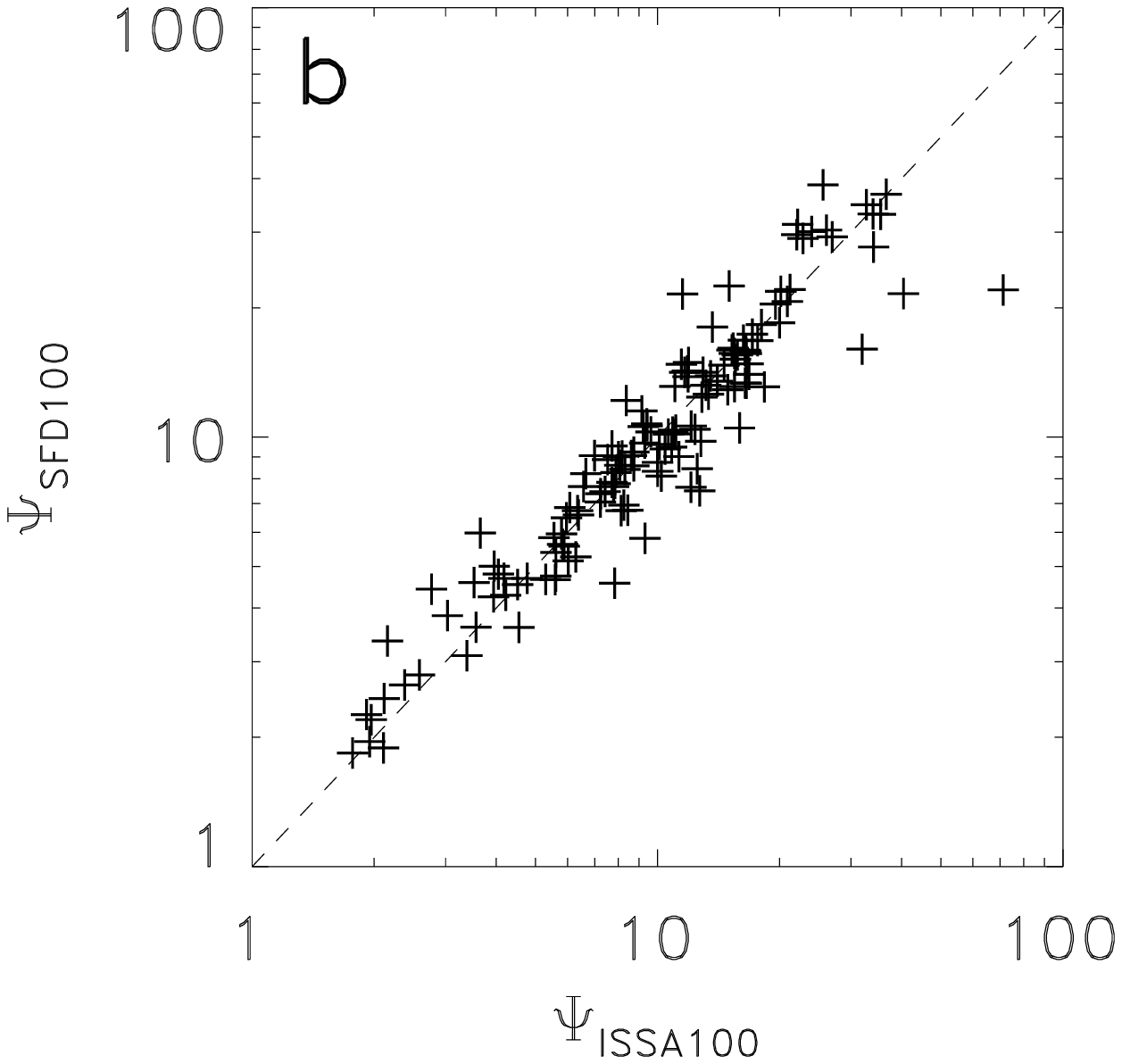, width=4.5cm}
\\
\vskip -0.3cm
\epsfig{file=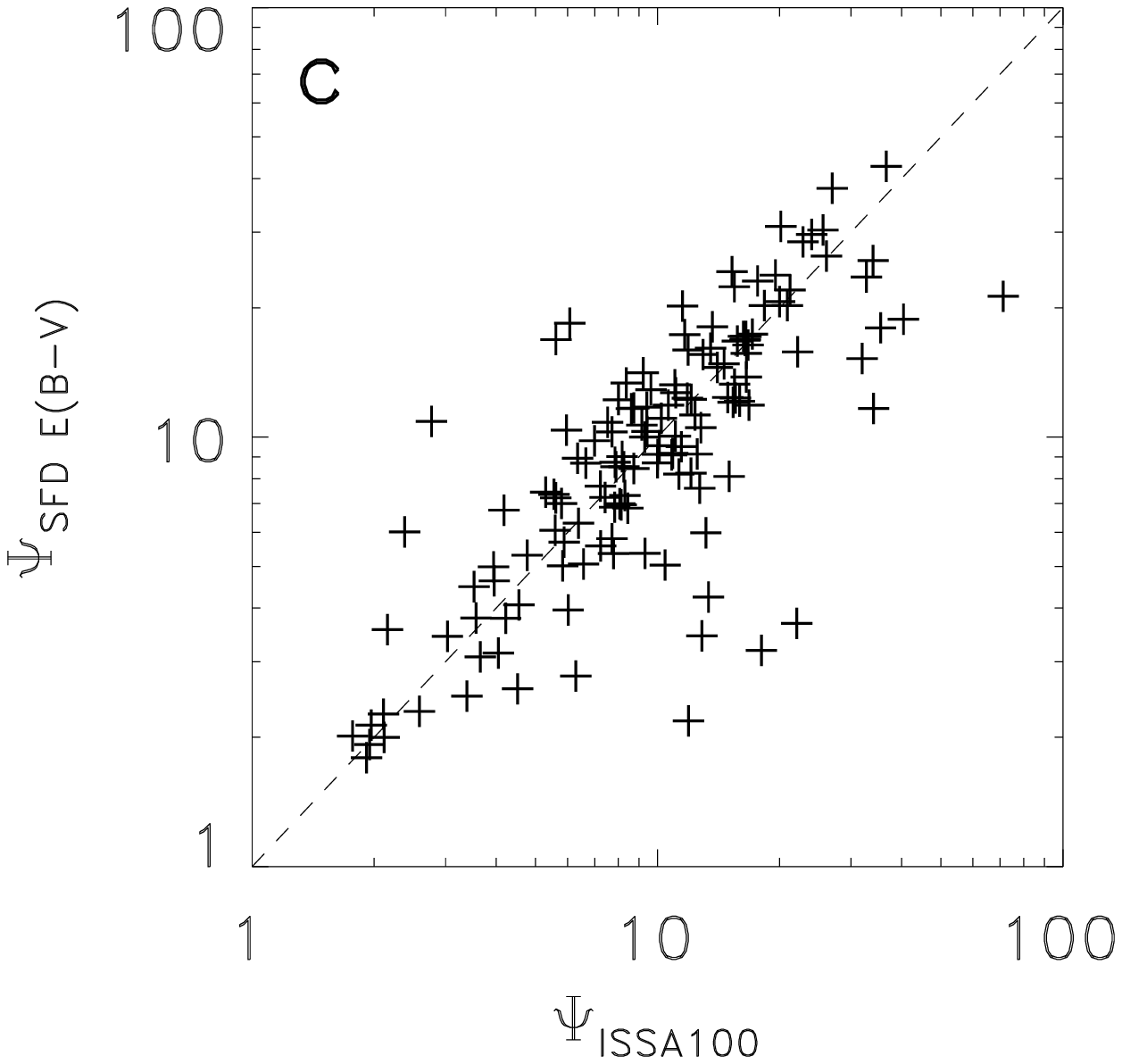, width=4.5cm}
\hskip -0.4cm \epsfig{file=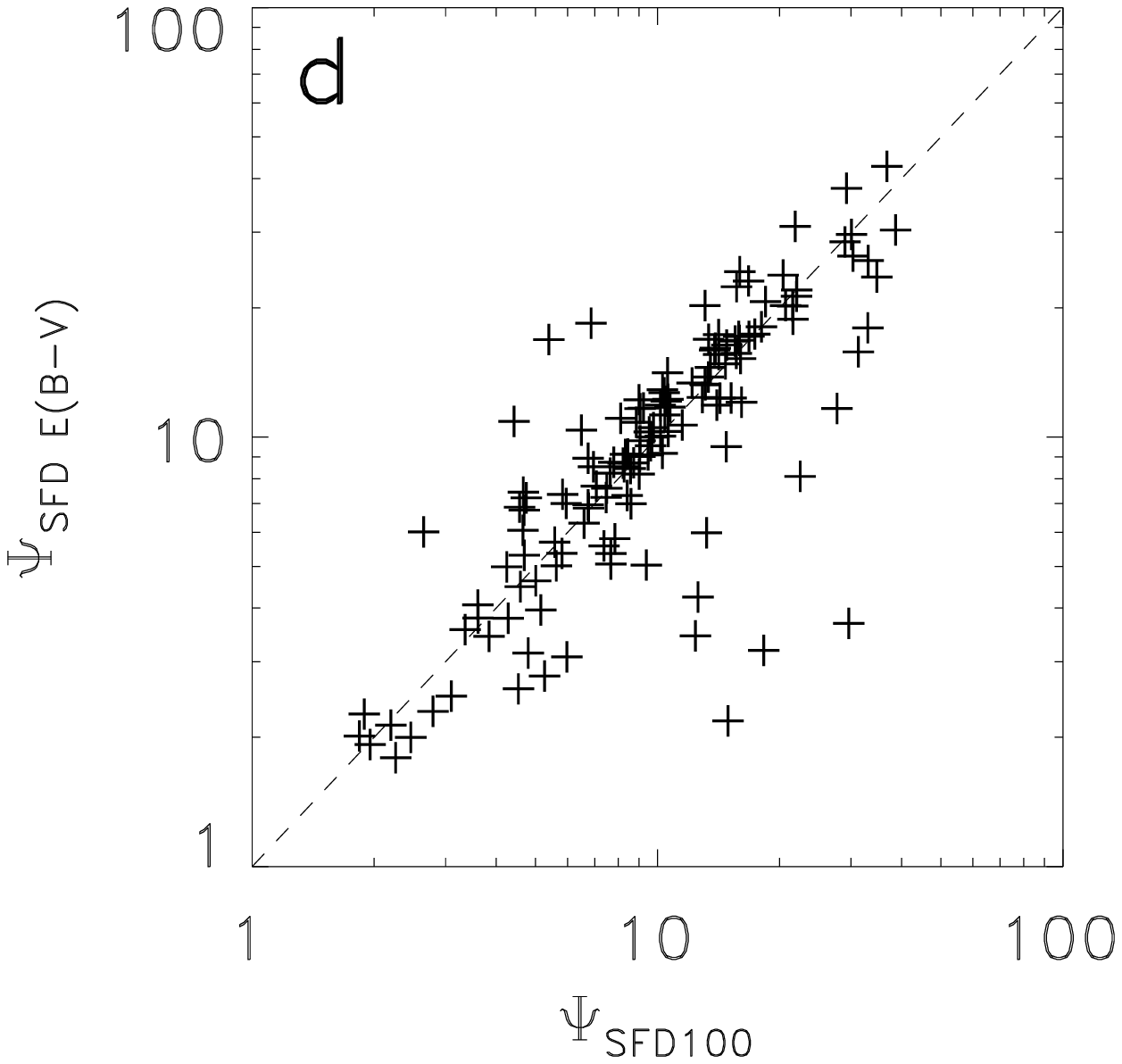, width=4.5cm}
\caption{Comparison of significances extracted from maps of 
different data sets. 
({\bf a}) ISSA 100\,$\mu$m vs. ISSA 60\,$\mu$m;
({\bf b}) ISSA 100\,$\mu$m vs. SFD 100\,$\mu$m;  
({\bf c}) ISSA 100\,$\mu$m vs. SFD reddening maps;
({\bf d}) SFD 100\,$\mu$m vs. reddening}
\label{fig:significances}
\end{figure}
%
\paragraph{Far-infrared colour of the loop walls:}
%
\begin{figure*}
\epsfig{file=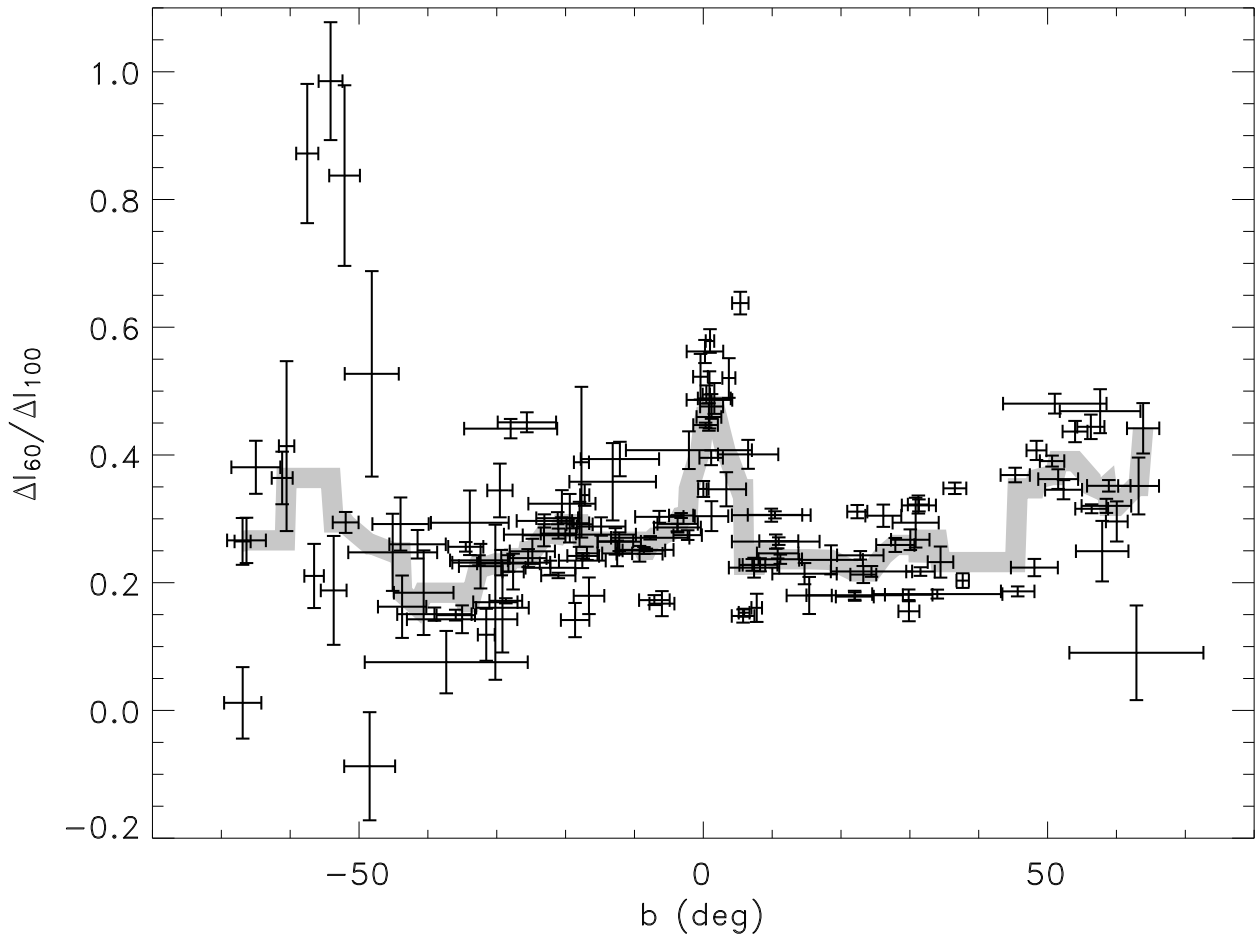, width=18cm}
%
\caption[]{Distribution of the colour indices of the loop walls 
($\rm\Delta I_{60}/\Delta I_{100}$) 
vs. galactic latitude. Error bars in galactic latitude 
represent the extension (r$_{eff}$) of the loop. Error bars of the 
colour indices are the formal errors of the I$_{100}$ vs.I$_{60}$
scatter plots. The thick gray curve represents the main trend.}
\label{fig:color}
\end{figure*}
As seen in Fig.~\ref{fig:color} most of the loops have a 
far-infrared colour similar to that of the extended galactic 
background emission or galactic cirrus with an average 
value of $\rm\Delta I_{60}/\Delta I_{100}$\,=\,0.25$\pm$0.12.
In the case of a few loops the far-infrared colour significantly deviates  
from that of the extended background, especially in the vicinity 
of the galactic plane ($\rm |b|\le10\degr$, 
see { Fig.~\ref{fig:color}}), where loops 
are "warmer" according to their FIR colours. 
A rather high scatter of FIR colours with high error bars can be observed 
for the highest galactic latitudes, which could also be related to the less
reliable accuracy of the surface brightness calibration in these fields. 
We conclude that the loops describe the structure of cirrus in the few to
$\sim$20\degr~ scales.


\begin{figure}
\epsfig{file=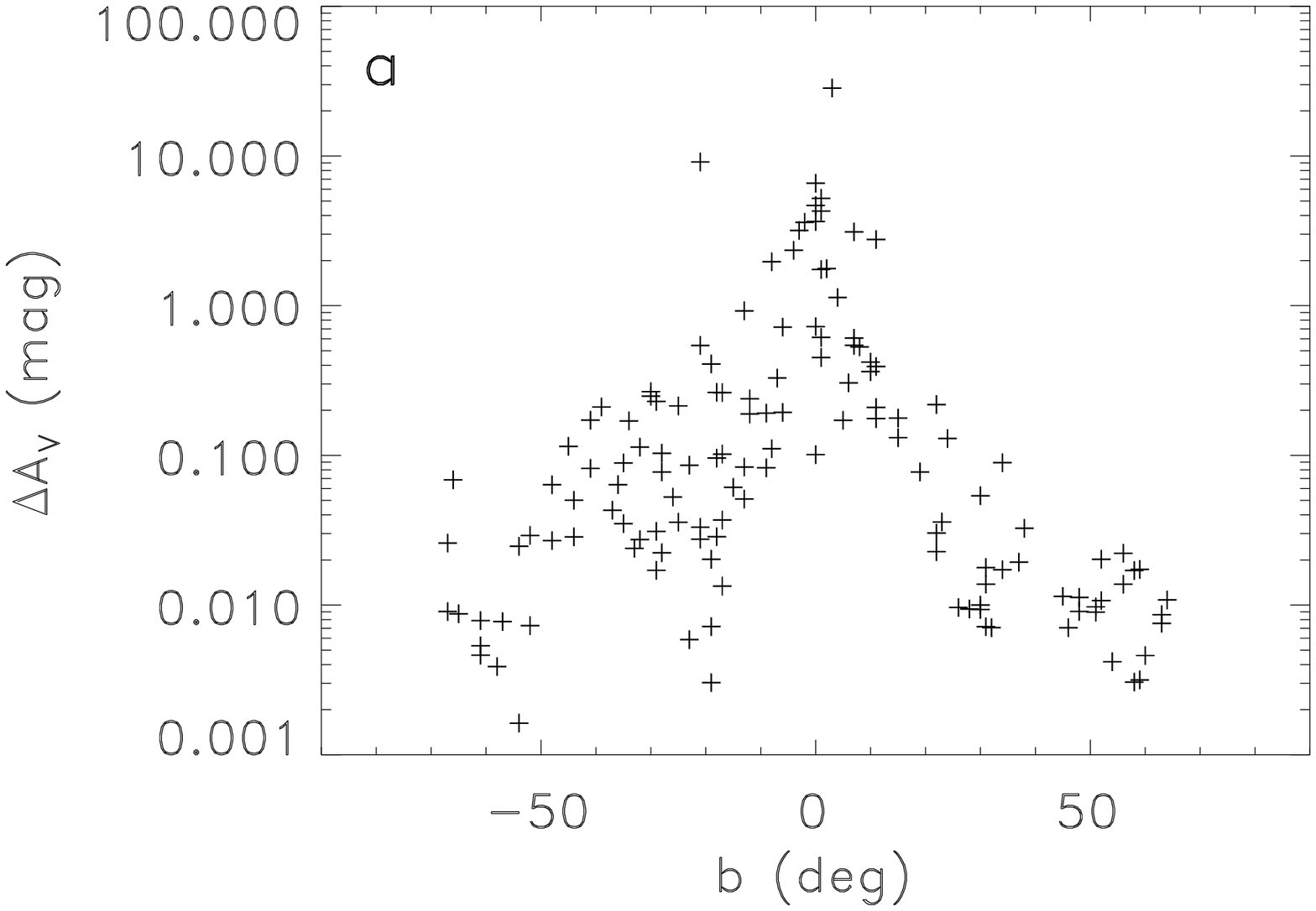, width=8.5cm}
\epsfig{file=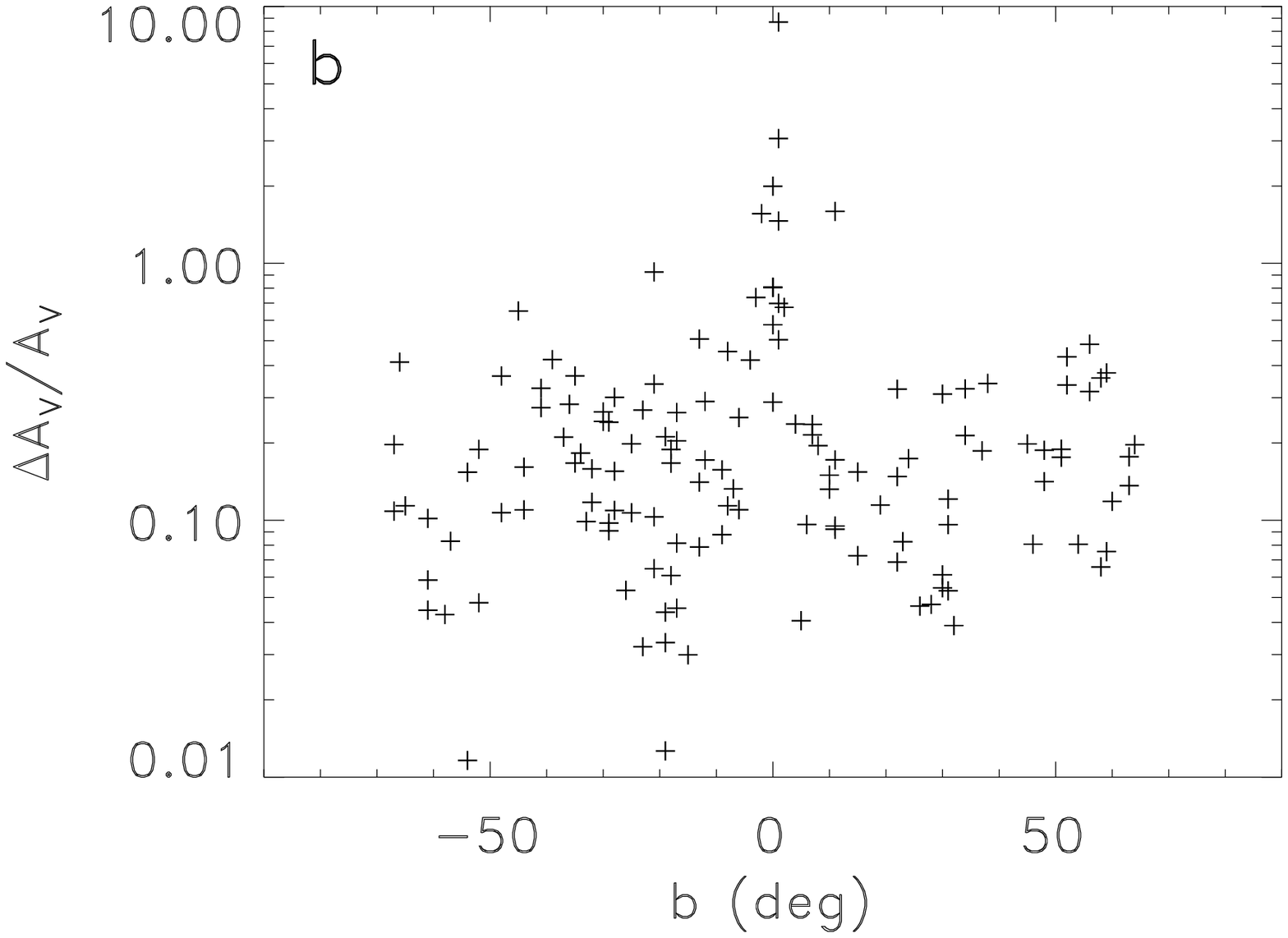, width=8.5cm}
\caption[]{Excess visual extinction in the loop walls derived 
from the SFD E(B-V) surface brightness profiles.
({\bf a}) galactic latitude vs. excess visual extinction $\Delta$A$_V$  
({\bf b}) galactic latitude vs. the ratio of the excess visual 
extinction $\Delta$A$_V$ to the average of the inner and outer visual 
extinction values ($\rm {A_V^{bg}} = {1\over{2}} [A_V^{in} + A_V^{out}]$) } 
\label{fig:AV}
\end{figure}
\paragraph{Excess extinction in the loop walls:}
Using the radially averaged SFD E(B-V) (reddening) profiles extracted from 
the appropriate maps we estimated the excess visual extinction
of the loop walls. Visual extinction values were calculated as
$\rm A_V = R \times E(B-V)$, where R is the ratio of total over
selective extinction and E(B-V) is the reddening value in a specific sky 
position. R was choosen to be uniform and set to the standard value of
3.1 of the diffuse medium \citep{Krugel}. 
The average value of A$_V$ was derived for the inner-, wall- and outer regions
(see Sect.~2.2), separately for all loops (A$_V^{in}$, A$_V^{wall}$ and A$_V^{out}$, 
respectively). 
We also derived an {\sl excess} visual extinction value for the 
loop wall, which was defined as 
\begin{equation}
\rm {{\Delta}A_V} = {{2\cdot{A_V^{wall}}}\over{A_V^{in}+A_V^{out}}}
\end{equation}
$\Delta$A$_V$ values show a clear dependence 
(despite the relatively large scatter around the main trend)
on the galactic latitude (Fig.~\ref{fig:AV}a), width values 
$<$0\fm01 at high galactic latitudes, while reaching 
$\Delta$A$_V$\,$\approx$\,10$^m$ in the vicinity of the galactic plane.

The relative strength of the excess extinction 
$\rm\Delta A_V$ relative to the background
extinction ($\rm A_V^{bg} = {1\over{2}}[A_V^{in} + A_V^{out}]$)
is usually low (in average $\sim$10\%, as 
shown in Fig.~\ref{fig:AV}b) except 
in the case of a few loops near the galactic plane. 
This corresponds to a $\sim$10\% increase of column density 
in the loop walls.  

\subsection{Distances to individual loops 
  \label{sect:distance}}

\paragraph{Derivation of the distance: \label{sect:dist_deriv}}
Many of the major physical properties of a loop (mass, 
size, particle number density, etc.) can only be tackled if the distance is
known. 
In order to estimate the distance of the loops, 
we used associated objects with known distances. 
There are two major classes of these
objects: (1) They may be responsible for the formation of the loop
itself, or are the remnants of the loop-creating events 
(we considered pulsars, luminous stars and/or their 
associations; see Sect.~\ref{sect:catalogue} for references). 
(2) They are part of the loop, but identified in another
wavelength range \citep[dark clouds and molecular clouds,][]{Dutra}. 

We applied the following main criteria when selecting distance indicators:

\begin{itemize}
  \item[i)] Luminous stars had to have the power to form the loop, 
    therefore only O stars and B stars with earlier subtype
     (or their associations) were taken into account. The sky position of the
     stars must be well confined in the central part of the loop 
     (although in some cases
     fast proper motions might have placed the star far from its 
      original position).
  \item[ii)] There must be at least two 
     objects (luminous star, pulsar, dark/molecular cloud) with 
     similar known distances associated with the loop, 
     otherwise the distance hypothesis would be rejected. 
  \item[iii)] The distances should fit each other within their 
  	uncertainties.     
\end{itemize}  

\begin{figure}[h!]
\epsfig{file=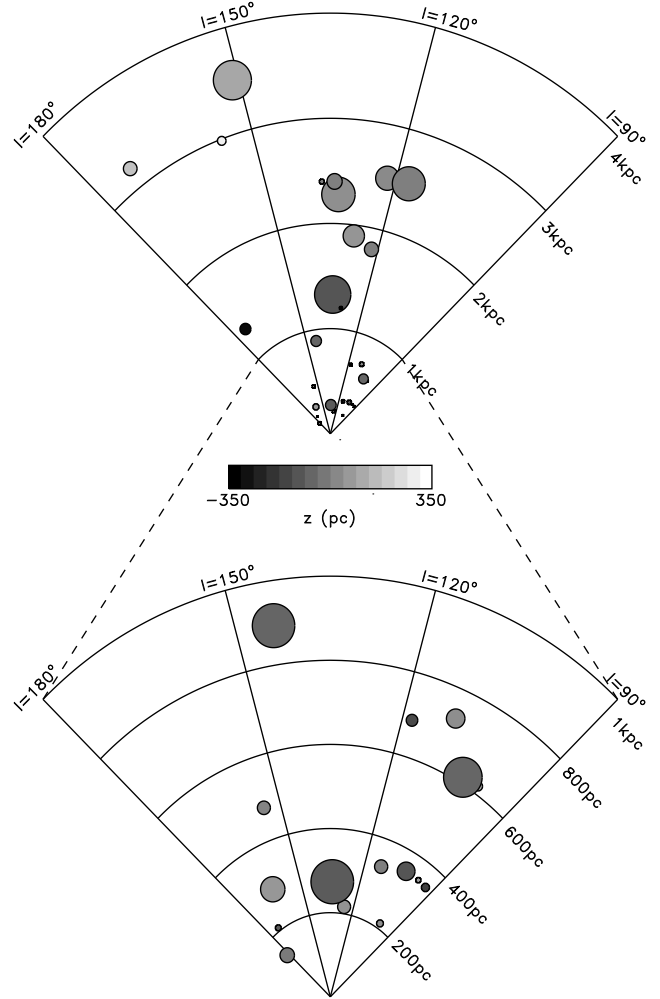, width=9cm}
\caption[]{Loops with known distances in the 2nd Galactic Quadrant,
projected to the galactic plane, in cylindriclateal coordinate system. 
The top panel shows the loops with projected distances up to 4\,kpc.
The bottom panel is a magnified view of the region closer than 1\,kpc. 
Different gray tones in both the upper and lower panels
code the vertical distance of the loop centres from the 
galactic plane (see colour bar). }
\label{fig:loop3d}
\end{figure}

\paragraph{Galactic distribution: \label{sect:dist_results}}
Based on the distances derived for some of our loops we drew the 
3-dimensional distribution
of GIRLs with known distances in Fig.~\ref{fig:loop3d}. 
\begin{figure}[h!]
\epsfig{file=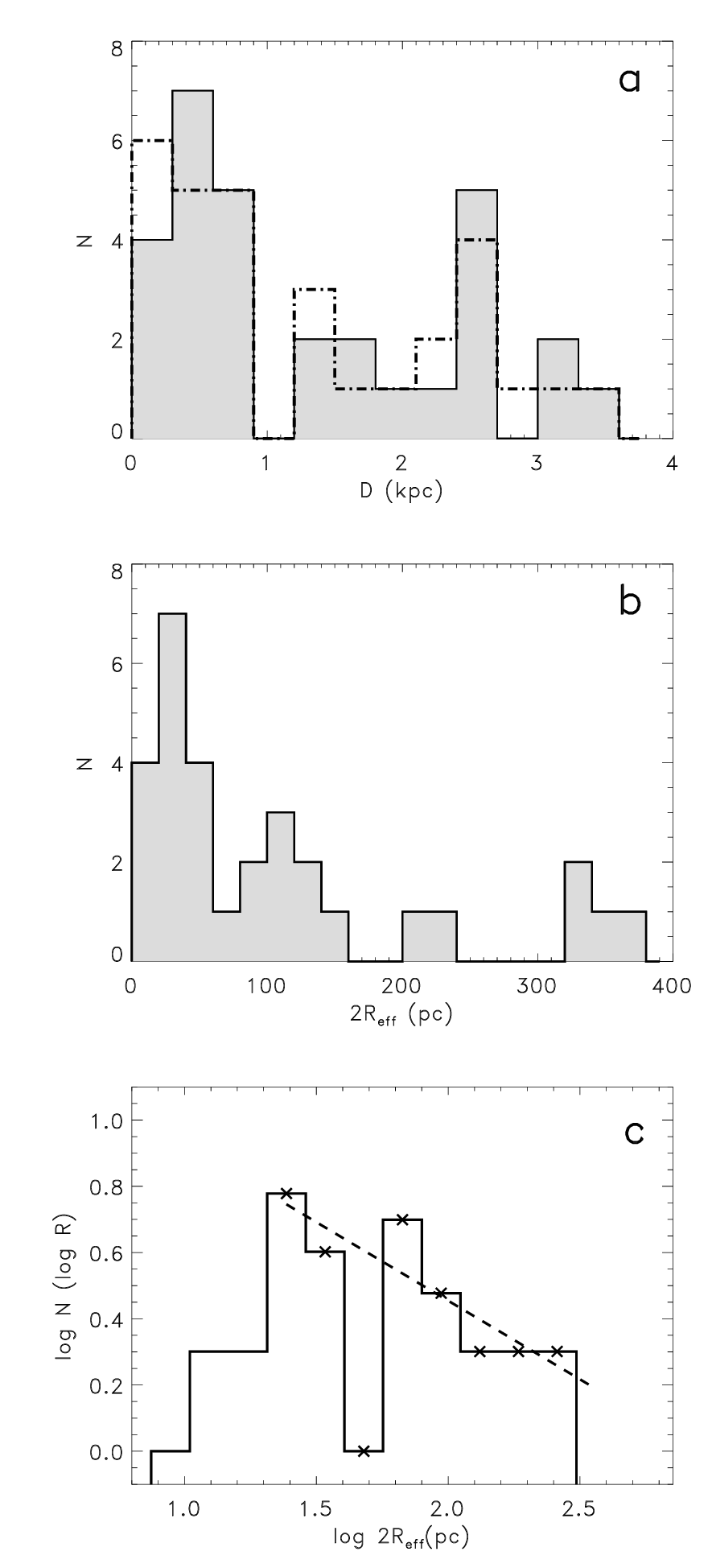, width=8.5cm}
\caption[]{({\bf a}) Distribution of distances among the loops presented
in Fig.\ref{fig:loop3d} (solid lines with gray filled area). 
Dashed-dotted lines represent the distribution of distances {\sl projected}
to the galactic plane. 
({\bf b}) Distribution of effective diameters (2$R_{eff}$) in the same sample. 
({\bf c}) The same as { b} but displayed and binned on a logarithmic scale. 
A relation of $\log 2R_{eff}$--$\log N$ was fitted for 
1.4\,$\le$\,$2R_{eff}$\,$\le$\,2.6. 
The points used for the fit are marked by asterisks
(see Sect.~4 for details).}
\label{fig:distance_hist}
\end{figure}

Since our distance indicators are mostly  
objects which are related to the spiral arms of the Galaxy,
the distribution reflects the position of the spiral arms,
introducing a selection effect. 
Loops closer than 800\,pc are related to OB associations and
luminous stars of the Local Arm and to the nearby interarm region, 
while another well-defined group
can be identified at a distance of 1.8--2.7\,kpc, in agreement with
the distance of the Perseus Arm in that direction
\citep{Humphreys}.
There is a hint that the majority of the GIRLs are within a distance
of $\sim$800\,pc. The most prominent loops remained observable up to
a distance of $\sim$3\,kpc. 



\section{Discussion \label{sect:discussion}}
%
%
\paragraph{Size distribution:}
In the Galaxy most of the ISM
is confined to a thin disc, forming the Galactic Disc itself
\citep[see e.g.][for a review]{Dickey+Lockman}.
Therefore it is expected that any object formed in the
ISM should somehow reflect the distribution of the originating medium.  
Accepting that our loops are features of the cold diffuse ISM, these
have been formed in the galactic disc with a scale height of $\sim$250\,pc. 
Therefore loops at high galactic latitudes
are close to our Solar System. On the other hand, in an exponential 
disc the number density of the gas and dust is decreasing rapidly with the
increasing height from the galactic plane. 
This allows the generation of larger SN- or stellar wind bubbles even 
possessing the same energy input as the ones closer to the galactic plane. 
Thus one would expect to observe loops with a large apparent size at high
galactic latitudes. Still, we found no correlation between the 
galactic latitude of the loop centres and the effective radii, neither on the
northern, nor on the southern galactic hemisphere. 

The sky distribution presented in Figs.~\ref{fig:mainsky} 
and \ref{fig:mainhistogram} 
shows clearly that the distribution of the GIRLs found at the 2$^{nd}$ 
Galactic Quadrant is far from that expected e.g. from the 
exponential disc hypothesis, in which the regions near to the
galactic plane should have the largest number of loops 
and this number should decrease drastically by the increasing 
galactic latitude. The relatively low number of loops 
in the galactic midplane may be explained by  multiple overlapping 
of the loop walls projected on the sky, i.e. many loops might be
undiscovered due to confusion. Intervening molecular clouds
may also contribute to this confusion. 

As discussed in Sect.~3.3, there are two separate distance 
groups in the known-distance sample. Accepting that the majority
of our loop are located in the nearby group (discussed 
in detail in the next paragraph) at a distance d\,$\le$\,800\,pc, 
one can obtain the expected ratio of loop counts above and below
$|b|=30\degr$ for a random distribution with a scale height of 
$z_H$\,=\,250\,pc \citep{Nakanishi}. This results in 
$(N_{low}/N_{high})_{calc}\,=\,$2.15 which should be compared 
with the observed ratio of $(N_{low}/N_{high})_{obs}\,=\,$1.28  
(without the contribution of the Perseus-Arm group). 
If this discrepancy is caused by confusion as discussed above, 
we may miss $\sim$40\% of the total count in the 
galactic midplane. 

It should be noted that 
molecular gas extends to galactic latitudes higher 
(both at the northern \citep[][fig. 2]{Hartmann98} and the
southern \citep[][fig. 1]{Magnani2000} galactic hemisphere) 
in the \second Galactic Quadrant than in the other parts of the sky.
This anomalous feature may also explain the richness of high galactic 
latitude loops in our sample. 
   
The large number of loops at high galactic latitudes 
might be a selection effect 
(loops are easier to be identified due to the lack of confusion).
Still, one needs a process that can generate loops
far from the galactic plane. One of these processes can be 
the infall of clouds from the halo, and their collision with the 
galactic disc. We know some examples, e.g. the NCP-loop  
\citep{Meyerdierks}, but the large number of such loops
would require a high rate of cloud infall, which is not
supported by observations \citep{Ehlerova}.

Another possibility is that turbulent processes in the ISM  create
'loop-like' structures \citep{Klessen,Korpi,Wada}.
Nonlinear development of turbulent instabilities 
may form voids in the ISM without the need of stellar energy injection.
These instabilities may be responsible for the large number of 
loop features at high galactic latitudes since these parts 
of the sky lack suitable stellar sources (massive O and/or B stars). 
From the observational point of view the 'loops' created by these
two different processes are practically indistinguishable, and 
therefore we are not able to determine the relative importance
of these processes.

It is also clear from the figures that local effects play 
a very important role forming the distribution. 
We can mention e.g. the high latitude group at 
l\,=\,130\degr--140\degr, b\,=\,45\degr--65\degr, which is
related to the huge NCP-loop and also belongs to a group 
of high latitude molecular clouds. Similar groups contribute much 
to the peaks in Fig.~\ref{fig:mainhistogram}a and b,
and might be responsible for the deviations from the expected
distributions. 
Since these structures are relatively nearby (e.g. the distance
of the NCP-loop is $\sim$330\,pc), even if we can detect loops
up to a distance of $\sim$3\,kpc,
the main contribution comes from loops at small distances
(see Sect.~3.3).


In { Fig.~\ref{fig:distance_hist}c} we displayed the size distribution on 
a logarithmic scale
and fitted the large diameter part (data points marked by asterisks) with a 
power-law (linear in the log--log plot). Similar investigation has been 
carried out by \citet{Oey} for the \object{Small Magellanic Cloud}, 
\object{Holmberg II}, \object{M\,31}, \object{M\,33} 
and by \citet{Kim2003} for the HI data of the Large Magellanic Cloud.  
We found a power law index of $s\,=\,-0.47\pm 0.11$ \citep[following][]{Kim2003}
or $\beta_0\,=\,1.24\pm 0.30$ using the transformation by \citet{Oey}, 
which is lower than that of the other galaxies mentioned above. 
We note that the distribution has a large scatter and it is better 
described by a double-peak one than by a power-law.

   
\paragraph{Correlation of distance and size:} 

\begin{figure}[h!]
\epsfig{file=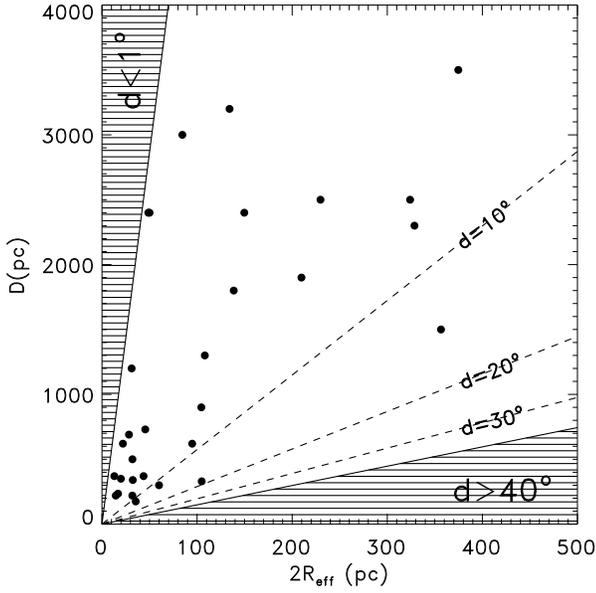, width=8.5cm}
\caption[]{Relationship between distance and effective diameter
(2$R_{eff}$) for loops with known distances (see Table~B.1). 
The shaded areas represent the apparent size limits 
2$r_{eff}$\,$\le$\,1$\degr$ and 
d\,$\ge$\,40$\degr$. Due to the limitations discussed in Sect.~2.1
we were not able to detect far-infrared features of these sizes.}
\label{fig:deff_vs_distance}
\end{figure}

A good correlation can be found between the physical
size of the bubble and the derived distance (for the sample in 
Sect.~3.3) when we derive 
linear Pearson correlation coefficients for the whole available 
distance range, up to 3.5\,kpc (0.66 and 0.73 for the linear-linear and
log-log plot, respectively).  Fig.~\ref{fig:deff_vs_distance} 
clearly illustrates the 
limitations of our study: due to the limitations in apparent size
(1\degr\,$\le$\,d\,$<$\,40\degr) we identified loops in the 
allowed region only. This selection effect is more severe for the
small-sized part of the distribution (1\degr\,$\le$\,2$r_{eff}$), while 
even the largest loop is a factor of $\sim$2 smaller than the
largest detectable feature. The distribution 
shows two groups, one at $D\le$800\,pc and one at 
$D>$800\,pc (Fig.~\ref{fig:distance_hist}a). 
Within the nearby group practically there is
{\sl no correlation} between the distance and physical size (with a linear
correlation coefficient of 0.17), therefore this group is not affected by
the selection effect. 
Accordingly, we may derive a statistical result for the local ISM, while
the Perseus Arm group is strongly biased. 
The size distribution of the nearby group lacks the large bubbles 
(2$R_{eff}$\,$>$\,100\,pc) which are observable
in the Perseus-group. 
The lack of large bubbles in the Local Arm group might be the consequence of
a non-continuous star-formation history in the Milky Way or
related to the finite extension of the Local Arm.   

\paragraph{Volume filling factor:}
The presence of the Perseus arm group is probably restricted to the
galactic plane and represent only a small subsample. Therefore it is 
plausible to assume that the majority of the loops with unknown distances
belong to the nearby (D\,$\le$\,800\,pc) group.    
Applying the distance -- size distribution obtained for the nearby group 
to all loops with unknown distances (except the loops of the Perseus group), 
it is possible to give a rough estimate for the volume filling factor 
\citep[or porosity Q,][]{Cox}
of our loop interiors in the galactic disc. Assuming that the 
loops/bubbles were formed
in the cold neutral interstellar medium (justified by their FIR colours) 
using $z_H$\,=\,250\,pc for the scale height of HI
\citep{Nakanishi} in the solar neighbourhood, we obtain a volume
filling factor of $f_{2^{nd}}$\,$\approx$\,4.6\% for the \second
Galactic Quadrant for D\,$\le$\,800\,pc. 
Taking into account an additional $\sim$40\% of loops in the midplane 
(see above) an upper limit of $f_{2^{nd}}$\,$\approx$\,6.4\% can be obtained.
These values confirm the filling factors by \citet{Ferrier98} 
and \citet{Gazol} 
($f$\,=\,3-8\% for 8.5\,$\le$\,$R_{\odot}$\,$\le$\,10\,kpc 
galactocentric distances).


\section{Summary and outlook \label{sect:summary}}
We summarize the main outcomes of our paper below:
\begin{itemize}
\item One fourth of the far-infrared sky was surveyed and
  145 loops were identified.
\item The loop sample is representative in the Local Arm
\item The FIR loops outside the midplane possess $\sim$10\%
  column density excess and a cirrus-like \irascolor\,$\approx$\,0.25
  ratio
\item The midplane loops hold several times denser walls and
  a far-infrared colour of \irascolor\,$>$\,0.4 
\item Distance estimates are given for 30 loops which trace the
  cold ISM in the Local Arm and in the Perseus Arm   
\item We estimated the hot gas volume filling factor to be
  4.6\%\,$\le$\,$f_{2^{nd}}$\,$<$\,6.4\% in the Local Arm in the \second 
  Galactic Quadrant   
\end{itemize}

As a continuation of this study the \first, \third
and \fourth galactic quadrants are surveyed as well. 
The complete sample and a comparison with HI distribution
will be given in a forthcoming paper.

\begin{acknowledgements}
\sloppy
This research was partly supported by the Hungarian Research Fund
(OTKA), grants \#F--022566 and \#T--043773.  
We are grateful to A. Burkert and B.G. Elmegreen for 
their valuable comments.
\end{acknowledgements}


\vskip 1cm
\large\bf
\noindent Appendix A, B and C are available at the following URL:
"http://astro.elte.hu/CFIRLG" as single PostScript files. 

\end{document}